# Monitoring LMXBs with the Faulkes Telescopes

**Fraser Lewis**[1]
*Faulkes Telescope Project / Las Cumbres Observatory Global Telescope Network (LCOGTN) / The Open University*
E-mail: `flewis@lcogt.net`

**David M. Russell**
*University of Amsterdam*
E-mail: `D.M.Russell@uva.nl`

**Rob P. Fender**
*University of Southampton*
E-mail: `r.fender@soton.ac.uk`

**Paul Roche**
*Faulkes Telescope Project / Las Cumbres Observatory Global Telescope Network (LCOGTN) / The Open University*
E-mail: `proche@lcogt.net`

**J. Simon Clark**
*The Open University*
E-mail: `S.Clark@open.ac.uk`

The Faulkes Telescope Project is an educational and research arm of the Las Cumbres Observatory Global Telescope Network (LCOGTN). It has two 2-metre robotic telescopes, located at Haleakala on Maui (FT North) and Siding Spring in Australia (FT South). It is planned for these telescopes to be complemented by a research network of eighteen 1-metre telescopes, along with an educational network of twenty-eight 0.4-metre telescopes, providing 24 hour coverage of both northern and southern hemispheres.

We have been conducting a monitoring project of 13 low-mass X-ray binaries (LMXBs) using FT North since early 2006. The introduction of FT South has allowed us to extend this to monitor a total of 30 LMXBs (see target list, Section 4). New instrumentation will allow us to expand this project to include both infrared wavelengths (z and y band) and spectroscopy. Brighter targets (~ 16 - 18 mag.) are imaged weekly in V, R and i' bands (SNR ~ 50), while fainter ones (> 18 mag.) are observed only in i' band (SNR ~ 20). We alter this cadence in response to our own analysis or Astronomers Telegrams (ATels).

---

[1] Speaker

## 1. Project Aims

1. To identify transient outbursts in LMXBs. LMXBs may brighten in the optical/near-infrared (OIR) for up to a month before X-ray detection. The behaviour of the optical rise is poorly understood, especially for black hole X-ray binaries. Catching outbursts from quiescence will allow us to examine this behaviour and alert the astronomical community to initiate multi-wavelength follow-up observations.

2. To study the variability in quiescence. Recent results have suggested that many processes contribute to the quiescent optical emission, including emission from the jets in black hole systems [1]. By monitoring the long-term variability of quiescent LMXBs, we will be able to provide some constraints on the emission processes and the mass fuctions.

## 2. GX 339-4

A target for FT South is the transient black hole binary GX339-4, which went into outburst in early 2007, followed by a steady decline in the following months. The outburst was detected at X-ray [2], OIR [3] and radio [4] wavelengths. Our observations (Figure 1) show that the source continued to decline in V, R and i' bands until ~ MJD 54585 (29 April, 2008) when the source increased in brightness to its previous brightest level. We have also noted rapid flaring in i' (Figure 2) and V bands with jumps of ~ 0.3 - 0.4 mags in periods of ~ 140 seconds [5]. Our ATel triggered multi-wavelength follow-ups with SALT, VLT, Swift and RXTE.

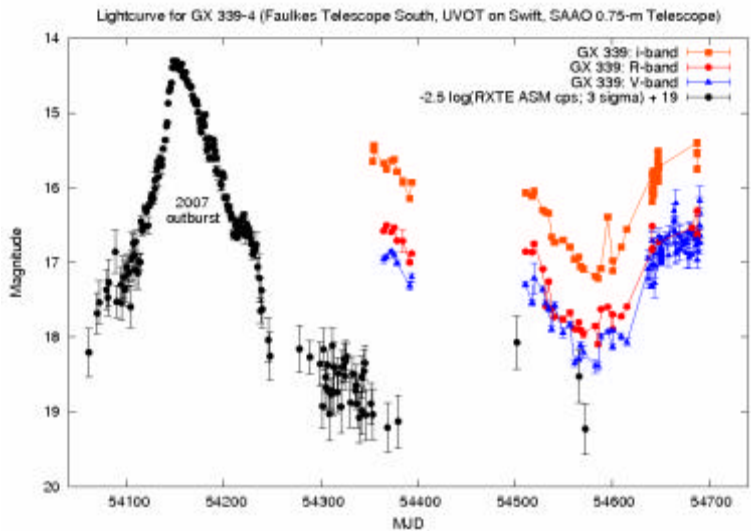

*Figure 1* FT ( orange, red, blue) and RXTE ASM data (black) of the 2007 outburst and 2008 re-flare of GX339-4. Swift UVOT V-band data are also included (blue from ~ 54650 – 54700).

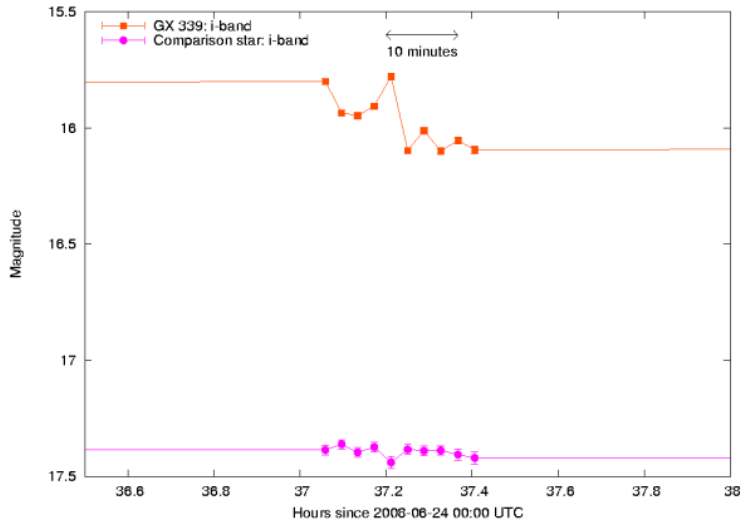

*Figure 2 Example of the rapid flaring behaviour seen in GX339-4*

### 3. XTE J1118+480

The bright Galactic Halo black hole system, XTE J1118+480, is monitored by several groups at different wavelengths [6] [7]. We note long-term variability over 800 days of observations. On occasion, we detect variability of ~ 0.6 mags in i'-band within a few days. In an attempt to discriminate between orbital modulation and flaring behaviour, we undertook observations in i'-band over a single orbit. This unfolded lightcurve (Figure 3) shows 0.4 mags variability in i' band over one orbit, which is known to be ~ 4.1 hours. [8]. This suggests that much of the long-term variability may be orbital in origin.

We intend to repeat these observations when the system is next visible in order to better constrain the ephemeris and to determine the orbital variability in V and R wavebands.

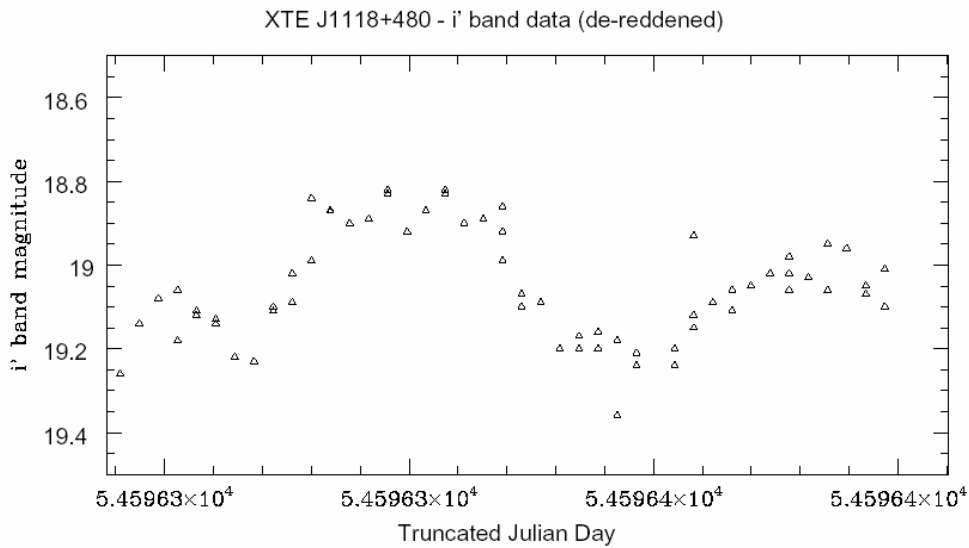

*Figure 3 FT unfolded i' band data of the black hole system, XTE J1118+480*

## 4. Targets

| System | Type | Orbital Period |
|---|---|---|
| IGR J00291+5934 (N) | MSXP | 2.46 hours |
| GRO J0422+32 (N) | BH | 5.092 hours |
| 4U 0614+09 (N) | Atoll, MQ | 50 minutes |
| A 0620-00 (N) | BH | 7.75 hours |
| XTE J0929-314 (S) | MSXP | 0.73 hours |
| GRS 1009-45 (S) | BH | 6.84 hours |
| XTE J1118+480 (N) | BH, MQ | 10.38 hours |
| GRS 1124-68 (S) | BH | 4.08 hours |
| GS 1354-64 (S) | BH, IMXB | 61.07 hours |
| Cen X-4 (S) | NS | 15.1 hours |
| 4U 1543-47 (S) | BHC, IMXB | 26.8 hours |
| XTE J1550-564 (S) | BH, MQ | 37.25 hours |
| 4U 1608-52 (S) | Atoll | 12.89 hours |
| 4U 1630-472 (S) | BH, MQ | ~ 1 day |
| XTE J1650-500 (S) | BH, MQ | 7.63 hours |
| GRO J1655-40 (S) | BH, IMXB, MQ | 62.88 hours |
| GX 339-4 (S) | BH, MQ | 42.14 hours |
| H 1705-250 (N) | BH | 12.51 hours |
| GRO J1719-24 (N) | BHC | 14.7 hours |
| XTE J17464-3213 (S) | BHC | none |
| SAX J1808.4-3658 (S) | MSXP | 2.014 hours |
| XTE J1814-338 (S) | MSXP | 4.275 hours |
| V4641 Sgr (S) | BHC | 2.8 days |
| XTE J1859+226 (N) | BHC, MQ | 9.16 hours |
| HETE J1900.1-2455 (S) | MSXP | 1.39 hours |
| Aql X-1 (N) | Atoll | 18.95 hours |
| 4U 1957+11 (N) | BHC | 9.33 hours |
| GS 2000+25 (N) | BH | 8.26 hours |
| V 404 Cyg (N) | BH | 155.4 hours |
| XTE J2123-058 (N) | Atoll | 5.96 hours |

N = FTN target

S = FTS target

Atoll = Atoll source neutron star LMXB

MSXP = Millisecond X-ray Pulsar

BH = Black Hole

BHC = Black Hole Candidate

IMXB = Intermediate-Mass X-ray Binary

MQ = displaying 'microquasar-type' behaviour

## 5. Online Microquasar Database

We are producing an interactive online resource of 'microquasar-type systems' with links to ADS, Simbad, etc. in a similar way to the open cluster database, WEBDA, or the galaxy database, NED. This will be an educational and research resource, which we will encourage researchers to contribute to.

It will contain fully-referenced data such as alternate names, luminosities, mass functions, distances, orbital periods, inclinations, finder charts and optical and X-ray lightcurves.

## 6. Acknowledgements

The Faulkes Telescope Project is an educational and research arm of the Las Cumbres Observatory Global Telescope Network (LCOGTN).

FL acknowledges support from the Dill Faulkes Educational Trust, and the support of the Institute of Physics' Research Student Conference Fund and the CR Barber Trust.

We acknowledge Craig Markwardt and the RXTE team for the PCA bulge scan data, which appears in Figure 2.

## 7. References

[1] Russell et al., *Global optical/infrared/X-ray correlations in X-ray binaries: quantifying disc and jet contributions*, MNRAS, **371** (2006) 1334.

[2] Krimm et al., *Swift-BAT detects a bright hard X-ray outburst from GX 339-4*, ATel (2006) 968.

[3] Buxton and Bailyn, *Latest optical and infrared observations of GX 339-4*, ATel (2007) 1027.

[4] Corbel et al., *ATCA radio observations of GX 339-4*, ATel (2007) 1007.

[5] Russell et al., *Unusual optical and X-ray flaring activity in GX 339-4*, ATel (2008) 1586.

[6] Gelino et al., *The inclination angle and mass of the black hole in XTE J1118+480*, ApJ, **642** (2006) 438.

[7] Zurita et al., *The 2005 outburst of the halo black hole X-ray transient XTE J1118+480*, ApJ, **644** (2006) 432.

[8] Patterson et al., *XTE J1118+480*, IAUC (2000) 7412.